\documentstyle[aps,prl,epsfig,multicol]{revtex}
\newcommand{\tr}{\mbox{Tr}\, }
\newcommand{\esc}{\gamma}

\renewcommand{\narrowtext}{\begin{multicols}{2} \global\columnwidth20.5pc}
\renewcommand{\widetext}{\end{multicols} \global\columnwidth42.5pc}
\newcommand{\Lrule}{\vspace*{-0.2in}\noindent\vrule width3.5in height.2pt
  depth.2pt \vrule depth0em height1em}
\newcommand{\Rrule}{\vspace{-0.1in}\hfill\vrule depth1em height0pt \vrule
  width3.5in height.2pt depth.2pt\vspace*{-0.1in}}

\begin{document}
\title{Noise through Quantum Pumps}
\author{M. L. Polianski$^a$, M. G. Vavilov$^{b,a}$, and P. W. Brouwer$^a$ }
\address{$^a$Laboratory of Atomic and Solid State Physics,
Cornell University, Ithaca, NY 14853}
\address{$^b$Theoretical Physics Institute, University of Minnesota,
Minneapolis, MN 55455}
\date{\today}

\maketitle
\begin{abstract}
We study the current noise through an unbiased quantum electron pump 
and its mesoscopic fluctuations for arbitrary temperatures and beyond 
the bilinear response. In the bilinear regime, we find the full 
distributions of the noise power and the current-to-noise ratio for 
a chaotic quantum dots with single-channel and many-channel 
ballistic point contacts. For a dot with many-channel point contacts 
we also calculate the ensemble-averaged noise at arbitrary 
temperature and pumping strength. In the limit of strong pumping,
a new temperature scale appears that corresponds
to the broadening of the electron distribution function in the dot as a
result of the time-dependent perturbations.\bigskip

\end{abstract}



\begin{multicols}{2}
\widetext
\narrowtext

\section{Introduction}

A periodic perturbation of a confined electron system may produce a 
direct current. The initial theoretical proposals \cite{thouless,Niu}
and experimental realizations \cite{Kouwenhoven,Pothier,Oosterkamp} 
for electron pumps were 
for a pump where the spectrum is gapped and the charge pumped in one 
cycle quantized and not subject to fluctuations. Recently, attention
has shifted to pumps that are well connected to electron reservoirs,
and, hence, do not have a gapped excitation spectrum 
\cite{FK,spivak,B_p,ZSA}. 
If the pump relies on a time-dependent perturbation that mainly
affects the quantummechanical
phases of the electrons, and not their classical trajectories, it is 
referred to as a ``quantum electron pump''. Such a
quantum pump was fabricated by Switkes {\em et al.} \cite{Switkes}. 

The current that is pumped through a quantum electron pump is subject 
to mesoscopic fluctuations and to quantum or thermal fluctuations
(noise).
Mesoscopic fluctuations of the current refer to the fact that the 
magnitude and
direction of the time-averaged current vary from sample to sample.
For a quantum pump built from a chaotic quantum dot, as is the
case in the experiment of Ref.\ \onlinecite{Switkes}, the 
mesoscopic current fluctuations were investigated
for various regimes of temperature,
pumping amplitude and dot conductances \cite{B_p,ZSA,SAA,Avron,VAA,Blaauboer}.
On the other hand, noise --- quantum and thermal fluctuations of the 
current --- is a property of the current pumped
through a particular realization of an electron pump. Noise in an
 electron pump is best
described by the fluctuations
of the charge pumped through the system in a certain number of
pumping cycles.
The statistics of such charge fluctuations was studied in Refs.\ 
\onlinecite{AK,L,AM,LL,LLL,IL,MM,buttikernew} for 
temperatures and pumping frequencies
much smaller than the inverse dwell time (escape rate) of electrons
in the quantum dot.

In this paper, we consider the mesoscopic fluctuations of the
noise. We do not impose any restrictions on the relative magnitudes
of temperature $T$, pumping frequency $\omega$, escape rate $\esc$
and pumping amplitude. This is important, as in an experimental
realization of a quantum pump, $T$ and $\esc$ are usually comparable,
while both are much larger than $\omega$. Previous works by Andreev
and Kamenev \cite{AK} and Levitov \cite{L} addressed the full counting
statistics, but at temperatures $k T \ll \hbar \omega$ only. The
mean square charge fluctuations for $\hbar \omega, k T \ll \esc$ (but 
including the case $\hbar \omega \approx k T$) were considered
very recently by Moskalets and B\"uttiker \cite{buttikernew}.

Denoting the quantummechanical average with a bar $\overline{\cdots}$,
the quantum and thermal fluctuations of the pumped charge are
described by
\begin{equation}
  S = {1 \over \tau_0}
  \left[ \overline Q^2 - (\overline Q)^2 \right].
  \label{eq:noise}
\end{equation}
Here $\tau_0$ is the observation time and
$Q$ is the total charge pumped through the dot in the time $\tau_0$.
The noise in an electron pump can be divided onto a 
Nyquist-Johnson component $S^N$ and pumping component $S^P$. 
The former is the thermal equilibrium noise due to the thermal
fluctuations of electrons in the leads, and depends on the 
electron temperature $T$ in the leads and 
the time-averaged conductance $\bar G$ of the quantum dot through
the fluctuation-dissipation theorem,
\begin{equation}
  S^N 
  = {2} k T \bar G.
  \label{eq:SN}
\end{equation}
The pumping contribution $S^P$, in contrast, is a true
non-equilibrium noise due to the perturbation of electrons {\it
inside} the dot. As we shall discuss in Sec.\ \ref{sec_2}, this
contribution to the noise can be seen as arising due to the
heating of electrons inside the dot as a result of the
time-dependent perturbation.

In the adiabatic regime $\hbar \omega \ll \gamma$ one needs to vary
at least two system parameters periodically in order to generate a
direct current. 
Current noise, however, is already generated if only one parameter is varied. 
The problem of current noise (and full counting statistics) for a 
single time-dependent scatterer with frequency $\hbar \omega \ll \gamma$ 
was addressed by Levitov and coworkers \cite{LLL,IL}. Our results can be
used to compute the mesoscopic fluctuations of the current in that
case. We find that
the main effect of the second time-dependent parameter
in a true electron pump is to reduce the mesoscopic fluctuations of
the noise.

The paper is organized as follows: A formal expression for the noise 
in terms of the time-dependent scattering matrix of the dot $\cal S$
is derived in Sec.\ \ref{sec_1}. 
Section \ref{sec_2} considers
the mesoscopic fluctuations of the noise through a chaotic quantum
dot with two (or more) time-dependent parameters. For an adiabatic
quantum pump at temperature $k T \ll \gamma$ we consider the full
distribution of the mesoscopic fluctuations of the noise in Sec.\
\ref{sec_21}. We focus on the cases of a quantum dot with
single-channel and many-channel point contacts. For a dot with
many-channel point contacts, the sample-to-sample fluctuations of
the noise are much smaller than the average.
In Sec. \ref{sec_22} we then present the ensemble-averaged
noise for a dot with many-channel point contacts at arbitrary 
temperature and pumping strength. We conclude in Sec.\ \ref{sec_3}. 
Finally, in Appendix we give 
some details of the calculations of the averages over the ensemble
of chaotic quantum dots.

\section{General formalism}\label{sec_1}

\begin{figure}
\centerline{\psfig{figure=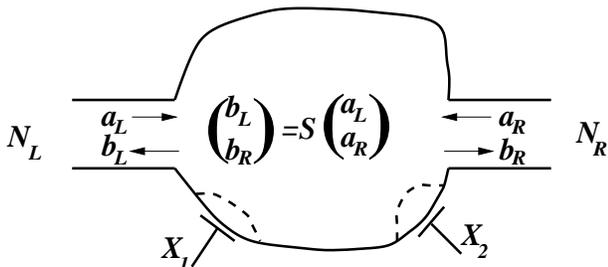,width=8cm}}
\caption {
Schematic drawing of the quantum dot and the leads. There are
$N_{L}$ ($N_{R}$) propagating modes in the left (right) contact. 
The shape of the dot is controlled by the voltages of two
shape-defining gates. The vectors ${\bf a}_{L,R}$ and ${\bf b}_{L,R}$
of annihilation operators for the incoming and outgoing  
states, respectively, in the left (L) and right (R) leads are 
related by the scattering matrix ${\cal S}$.
}
\label{fig:stub}
\end{figure}

We consider fluctuations of charge transmitted the sample during 
the observation time $\tau_0$. The sample is connected to electron
reservoirs through two point contacts with $N_{L}$ and $N_{R}$ open 
channels, respectively, see Fig.\ \ref{fig:stub}. 
It is subject to a periodic
perturbation at frequency $\omega/2\pi$; The perturbation is
described by specifying the time-dependence of 
parameters $X_1(t)$, $X_2(t)$, \ldots, $X_n(t)$ characterizing
Hamiltonian of the sample. The electrons in the two
reservoirs are held at the same chemical potential $\mu$ and 
temperature $T$ at all times during the pumping cycle. 

We calculate the noise of the quantum pump using the scattering
formalism of B\"uttiker \cite{scatt}. In B\"uttiker's original 
application
there are no time-dependent perturbations, so that the system is 
described by a scattering matrix that depends on energy, but not
on time. On the other hand,
as long as pumping frequencies and temperatures much 
smaller than the escape rate $\esc$ are considered, the system
with a time-dependent perturbation can be described using a
scattering matrix that depends on time, but not on energy 
\cite{AK,L,LLL,IL,MM,buttikernew}. 
When $k T$, $\hbar \omega$ and $\gamma$
are all comparable, one needs to use a scattering matrix ${\cal S}$
that depends on both time and energy, or, equivalently, that 
depends on two times or two energies. 
Here we shall make use of a formulation with a scattering
matrix ${\cal S}(t,t')$ that depends on two times. This
formulation was used to calculate
the the time-averaged conductance and pumped current in the Ref. \cite{VAA}. 
The formalism is equivalent to the two-energy scattering
formalism developed by B\"uttiker and coworkers for time-dependent
transport through mesoscopic 
structures \cite{buttiker1,buttiker2,BC,BB}.

The scattering matrix ${\cal S}(t,t')$ relates the annihilation
[creation] operators ${\bf a}_{\alpha}(t)$ [${\bf a}^{\dagger}(t)$] and 
${\bf b}_{\alpha}(t)$ [${\bf b}^{\dagger}_{\alpha}(t)$] of incoming states and
outgoing states in channel $\alpha=1,\ldots,2N$ of the leads (the
index $\alpha$ includes the spin degree of freedom),
\begin{mathletters}
\label{modes}
\begin{eqnarray}
{\bf b}_{\alpha}(t)&=&
  \int_{-\infty}^{+\infty} {\cal S}_{\alpha\beta}(t,t')
  {\bf a}_{\beta}(t')dt',\\
  {\bf b}_{\alpha}^{\dagger}(t)&=&\int_{-\infty}^{+\infty}
  {\bf a}_{\beta}^\dagger(t')
  ({\cal S}^\dagger(t',t))_{\beta\alpha}dt'.
\end{eqnarray}
\end{mathletters}
Here the indices $\alpha$ and $\beta$ label
the propagating channels in the point
contact contacting the dot to the left and right reservoirs for 
$\alpha,\beta=1,\ldots,2N_{L}$ and $\alpha,\beta=2N_{L}+1,\ldots,2N$, 
respectively.
The $2N \times 2N$ matrices ${\cal S}^\dagger$ and ${\cal S}$ are
related as
\begin{eqnarray}
\left({\cal S}^\dagger(t',t)\right)_{\alpha\beta}=
{\cal S}_{\beta\alpha}^*(t,t').
\end{eqnarray}
Causality requires that ${\cal S}(t,t') = 0$ if $t < t'$. We restrict
our attention to the case where spin rotation invariance is preserved,
and, hence, ${\cal S}$ is proportional to the $2 \times 2$ unit matrix
in spin grading.

The expression for the current ${\bf I}_{L}$ in the left lead is 
\begin{eqnarray}
\label{currents}
  {\bf I}_{L}(t) &=& e
  \sum_{\alpha=1}^{2N_{L}} 
  \left[{\bf a}_{\alpha}^\dagger(t) {\bf a}_{\alpha}(t) -
  {\bf b}_{\alpha}^\dagger(t){\bf b}_{\alpha}(t)\right].
\end{eqnarray}
A similar expression holds for the current ${\bf I}_{R}$ in the right
lead. Since we are only interested
in the charge pumped through the dot in a long time interval and the
charge on the dot is conserved after each cycle, we replace the
expression for the current operator by a suitable combination of ${\bf
I}_{L}$ and ${\bf I}_{R}$,
\begin{mathletters}
\begin{eqnarray}
\label{current}
  {\bf I}(t) &=& {N_{R} \over N} {\bf I}_{L}(t) - {N_{L} \over N} 
  {\bf I}_{R}(t) \nonumber \\ &=&
  e \sum_{\alpha,\beta=1}^{2N} 
  \left({\bf a}_{\alpha}^\dagger(t){\Lambda}_{\alpha\beta}
  {\bf a}_{\beta}(t)-
  {\bf b}_{\alpha}^\dagger(t)\Lambda_{\alpha\beta} 
  {\bf b}_{\beta}(t)\right).
\end{eqnarray}
Here $\Lambda$ is a diagonal matrix with elements
\begin{equation}
  \Lambda_{\alpha\alpha} = \left\{ \begin{array}{ll} N_{R}/N & 
\alpha=1,\ldots,2N_{L}, \\
  -N_{L}/N & \alpha=2N_{L}+1,\ldots,2N. \end{array} \right.
\end{equation}
\end{mathletters}

In terms of the current operator ${\bf I}$, the time-averaged 
pumped current $I$ reads
\begin{equation}
  I = {1 \over \tau_0}
  \int_0^{\tau_0} dt
  \overline{\bf{I}}(t),
\end{equation}
where the observation time $\tau_0$ is an integer number of
pumping cycles.
The noise $S$ is defined as [cf.\ Eq.\ (\ref{eq:noise})]
\begin{equation}
  S = {1 \over \tau_0}
  \int_0^{\tau_0} dt dt'
  \left( \overline{\bf{I}(t) \bf{I}(t')} - 
   \overline{\bf{I}}(t) \overline{\bf{I}}(t')\right).
\end{equation}

In the leads, the electron distribution function is given
by the Fourier transform $f(t)$ of the Fermi function,
\begin{eqnarray}
  \overline{{\bf a}_{\beta}^\dagger(t'){\bf a}_{\alpha}^{\vphantom{\dagger}}(t)}
  &=& \delta_{\alpha \beta} f(t'-t), \nonumber \\
  \overline{{\bf a}_{\beta}^{\vphantom{\dagger}}(t'){\bf a}_{\alpha}^{\dagger}(t)}
  &=& \delta_{\alpha \beta} \tilde f(t-t'),
  \label{eq:aavg}
\end{eqnarray}
where we defined
\begin{equation}
  \tilde f(t) = \delta(t) - f(t).
\end{equation}
Using Eq.\ (\ref{modes}) to eliminate the operators 
${\bf b}(t)$ and ${\bf b}^{\dagger}(t)$ from Eq.\ (\ref{current})
and Eq.\ (\ref{eq:aavg}) to compute the quantummechanical
expectation values, we find the average current
\begin{eqnarray}
\nonumber
  I &=& 
  {1 \over \tau_0} e\int_{0}^{\tau_0}dt\int dt_1dt_2 f(t_1-t_2)\\
  && \times
  \tr\left[
  {\cal S}^\dagger(t_1,t)\Lambda{\cal S}(t,t_2)-\delta(t-t_1)\Lambda
  \delta(t-t_2)\right],
  \label{q1}
\end{eqnarray}
and the noise
\begin{eqnarray}
  S &=& S^N + S^P,
\end{eqnarray}
where the Nyquist-Johnson and pumping contributions $S^N$ and $S^P$ read
\widetext
\Lrule
\begin{eqnarray}
  S^N &=& {2 e^2 \over \tau_0}
  \int_{0}^{\tau_0} dt dt'\int dt_1 dt_2
  f(t_1-t')\tilde f(t'-t_2)\,
  \tr\left[ \delta(t-t_1) \Lambda^2 \delta(t-t_2)-
  {\cal S}^\dagger (t_1,t)\Lambda {\cal S}(t,t_2)\Lambda
  \right], \\
  S^P &=& {e^2 \over \tau_0}
  \int_{0}^{\tau_0} dt dt'\int dt_1 dt_2 dt'_1 dt'_2
  f(t_1-t'_2)\tilde f(t'_1-t_2) \nonumber \\ && \mbox{} \times
  \tr\left[
  {\cal S}^\dagger(t_1,t)  \Lambda {\cal S}(t,t_2)
  {\cal S}^\dagger(t'_1,t')  \Lambda {\cal S}(t',t'_2)
  -
  \delta(t-t_1)\delta(t'-t_1')\Lambda^2\delta(t-t_2)
  \delta(t'-t_2')\right].
  \label{q2_pumping}
\end{eqnarray}
\Rrule\narrowtext\noindent
(These equations can also be derived using the Keldysh formalism, see
Ref.\ \onlinecite{mvthesis}.)

For adiabatic pumping, $\hbar \omega \ll \esc$, Equation (\ref{q1})
is equivalent to the time-averaged current of Refs.\ \onlinecite{B_p}
and \onlinecite{ZSA}. The Nyquist-Johnson contribution to the
noise is related to the time-averaged conductance $\bar G$
of the system at temperature $T$, see Refs.\ 
\onlinecite{conductance,transport}
and Eq.\ (\ref{eq:SN}) above.

\subsection{Bilinear adiabatic pumping}\label{sec_11}

Of particular interest is the case when the perturbation is slow
compared to the (elastic) escape rate $\gamma$
of the electrons from the
sample into the reservoirs. This is the regime of the adiabatic
quantum pump of Ref.\ \onlinecite{Switkes}.
In this approximation it is advantageous to use an 
analog of the Wigner transform for the matrix ${\cal S}(t,t')$:
\begin{eqnarray}\label{wigner}
{\cal S}(\varepsilon, t)=\int_{-\infty}^{+\infty}dt' e^{-i\varepsilon(t-t')}
{\cal S}(t,t').
\end{eqnarray}
Up to corrections of order $\hbar \omega/\gamma$, the 
matrix ${\cal S}(\varepsilon,t)$ is equal to the ``instantaneous''
scattering matrix ${\cal S}_X(\varepsilon)$, which is obtained by
``freezing'' all parameters $X_j$ to their values at time $t$
\cite{BL,AppC}.

If, in addition to being adiabatic, the parameters $X_j$, 
$j=1,\ldots,n$, undergo only small excursions from their
average value, which we set to zero, we may further expand in $X_j$.
To find the noise it is sufficient to expand ${\cal S}$ up to the 
second order in $X$. Arranging the parameters $X_j$ in an
$n$-component vector ${\cal X}=(X_1,\ldots,X_n)^{\rm T}$, we thus find
\widetext
\Lrule
\begin{mathletters} \label{bilinear}
\begin{eqnarray}
\nonumber
S^P & = & {e^2 \over 4 \tau_0} \int_0^{\tau_0} dtdt' 
\int_{-\infty}^{+\infty}
\frac{d\varepsilon}{2\pi}\int_{-\infty}^{+\infty}\frac{d\varepsilon'}{2\pi}
\cos\left[(\varepsilon-\varepsilon')(t-t')\right]
\left[f(\varepsilon) \tilde f(\varepsilon')+f(\varepsilon)\tilde f(\varepsilon')
\right]
\\
&& \mbox{} \times
  \left[({\cal X}^{\rm T}(t) - {\cal X}^{\rm T}(t'))
   {\cal K}(\varepsilon,\varepsilon')
  ({\cal X}(t')- {\cal X}(t))\right],
\label{Q_general}
\end{eqnarray}
where $\tilde f(\varepsilon) = 1 - f(\varepsilon)$ and
the $n \times n$ matrix ${\cal K}$ reads
\begin{eqnarray}
\label{Kij_general}
{ \cal K}_{ij}(\varepsilon,\varepsilon') & = &
\tr \left[ {\Lambda}^2 {\cal R}_i(\varepsilon) {\cal R}_j(\varepsilon)+
{\Lambda}^2 {\cal R}_j(\varepsilon'){\cal R}_i(\varepsilon')
-2\Lambda {\cal R}_i(\varepsilon) \Lambda
{\cal R}_j(\varepsilon')
\right], \ \ \label{R}
{\cal R}_j(\varepsilon)=-i\frac{\partial{\cal S}_X(\varepsilon)}{\partial X_j}
{\cal S}_X^\dagger(\varepsilon).
\end{eqnarray}
\end{mathletters}
\Rrule
\narrowtext

In the bilinear regime and at zero temperature, the 
time-averaged current $I$ and the noise
$S^P$ are sufficient to define the total counting statistics of
the pump \cite{L}: To lowest order in the excursions of the 
parameters $X_j$, the pumping cycles are statistically independent.
A pumping cycle is characterized by quantummechanical
probabilities $P_{R}$ and $P_{L}$ that an electron is pumped from left 
to right or from right to left, respectively. Both $P_{R}$ and $P_{L}$ 
are small, of order $X^2$, and one has $I = (e \omega/2\pi)(P_{L}-P_{R})$, 
$S^P = (e^2 \omega/2\pi)(P_{L}+P_{R})$.

For a typical quantum dot with capacitance $C$ and mean level
spacing $\Delta$, the charging energy $e^2/2 C\gg\Delta$. The
derivatives in Eq.\ (\ref{R}) should be taken at a constant
value of the chemical potential $\mu$, which, in the Hartree
approximation, is equal to sum of 
the electron's kinetic energy and
the electrostatic potential. In the absence of electron-electron
interactions all derivatives are taken at constant value of the
kinetic energy.
As we prefer to take derivatives at constant kinetic energy
(i.e., at constant $\varepsilon$) in both cases, we 
substitute the parametric derivatives as \cite{buttiker1,parametric}.
\begin{eqnarray}\label{eq:vgandc}
\left. \frac{\partial}{\partial X} \right|_{\mu}
 &\to & \left. \frac{\partial}{\partial X} \right|_{\varepsilon}
\mbox{} - \left(
\frac{{1 \over 2}\tr {\cal R}_i}
{{\pi C \over e^2} - {i \over 2} \tr {\partial {\cal S}_X \over \partial
 \varepsilon} {\cal S}_X^{\dagger}}
 \right) {\partial \over \partial \varepsilon}.\label{derivative}
\end{eqnarray}
Equation (\ref{derivative}) differs from a similar expression in 
Ref. \cite{B_p} by factor of $1/2$ in front of the traces because
of the double size of matrix $\cal S$ as a result of the inclusion
of spin.
\section{Application to chaotic quantum dots}\label{sec_2}

We now consider the mesoscopic fluctuations of the pumping noise
$S^P$ for the case of a chaotic quantum dot. The quantum dot
is characterized by a mean level spacing $\Delta$, escape rate
$\gamma = N \Delta/2 \pi$, Thouless energy
$E_{\rm Th} = \hbar/\tau_{\rm erg} \gg \Delta$, and capacitance
$C$, with charging energy $e^2/2 C \gg \Delta$.
It is coupled to two electron
reservoirs via ballistic point contacts with $N_{L}$ and $N_{R}$
channels each, see Fig.\ \ref{fig:stub}. 

The dot is driven by periodically varying parameter(s)
$X_i$, $i=1,\ldots,n$ with frequency $\omega$,
cf. Fig. \ref{fig:stub}. In the
experiment of Ref.\ \onlinecite{Switkes}, the parameters $X_i$ 
correspond to the voltages on external gates that control the dot
shape. The precise relation between the parameters $X$ used in the
theory and in the experiment is not known a priori, but can be
established using independent measurements of, e.g., the derivative
of the conductance \cite{parametric}
or the rate of change of the position of Coulomb
blockade peaks when the point contacts between the dot and the
reservoirs are pinched off \cite{VAA}. Following Refs.\
\onlinecite{B_p,SAA,VAA}, we will assume that the
different parameters $X_i$ correspond to different perturbations
of which the matrix elements between states (of the closed dot)
within a Thouless energy from the Fermi level $\varepsilon=0$
are Gaussian and independently distributed.
We choose the scale for the parameters $X_i$ such that the
mean square derivative $\langle (\partial \varepsilon_{\mu}/
\partial X_i)(\partial \varepsilon_{\mu}/
\partial X_j) \rangle = \delta_{ij}\Delta^2/\pi^2$, where $\Delta$ is the mean
level spacing and $\varepsilon_{\mu}$ is an energy level in the
closed dot.

The transmitted charge $Q$ is measured during a time $\tau_0$, which
we will assume to be a large number of pumping cycles.
This requirement of large observations is discussed in detail
in Ref.~\cite{IL}. For short observation times boundary effects 
related to switching processes in the system need to be taken into
account.
\subsection{Weak adiabatic low temperature pumping}
\label{sec_21}

If not only the frequency $\hbar \omega$ is much smaller than the 
escape rate $\esc$, but also $k T \ll \esc$, the scattering matrix 
$S(\varepsilon,t)$ in Eq.\ (\ref{bilinear}) can be taken at
the Fermi level $\varepsilon=0$, and we find the simple result
\begin{mathletters}
\begin{eqnarray}
\label{bilinear1}
S^P & = &\frac{e^2}{\tau_0}\int_0^{\tau_0}dtdt'f(t-t')\tilde f(t'-t)
\nonumber \\
&\times & \frac 12\left[({\cal X}(t)^{\rm T}- {\cal X}(t')^{\rm T})
{\cal K} ({\cal X}(t')-{\cal X}(t))\right], \\
\label{Kij}
{ \cal K}_{ij} & = &
\tr \left( {\Lambda}^2 {\cal R}_i {\cal R}_j+{\Lambda}^2
{\cal R}_j {\cal R}_i-2 \Lambda {\cal R}_i \Lambda {\cal R}_j
\right) .
\end{eqnarray}
\end{mathletters}

We now consider the case of two time-dependent parameters,
${\cal X}(t)=(X_1(t),X_2(t))^{\rm T}$:
\begin{equation}
X_1(t)=X_1 \cos(\omega t),\ X_2(t)=X_2 \cos(\omega t + \phi),
  \label{eq:X}
\end{equation}
in more detail. For this case, Eq. (\ref{bilinear1}) 
is factorized into a factor $F(T,\omega,\tau_0)$ that
depends on the relevant time and energy scales, and a sample
specific contribution,
\begin{eqnarray}
  S^P  &=&e^2 F(T,\omega,\tau_0)\left({\cal K}_{11}X_1^2+{\cal K}_{22}X_2^2
\right.\nonumber \\ && \left. \mbox{} + 2\cos\phi\ {\cal K}_{12}X_1^{\vphantom{1000}}X_2^{\vphantom{1000}}\right).
\label{vector}
\label{factor}
\end{eqnarray}
To find the integral $F(T,\omega,\tau_0)$ for the physically
relevant limit of long observation times $\tau_0 \gg 1/T$, we
note that in this limit the distribution function $f(t)$ is given 
by Fourier transform of the equilibrium Fermi distribution function,
\begin{eqnarray}
\label{fdistribution}
f(t)=\int {d \varepsilon \over 2 \pi \hbar}
 \frac{e^{i\varepsilon t/\hbar}}
{e^{\varepsilon/kT}+1}=\frac{ikT}{2 \hbar \sinh(\pi k T t/\hbar)}.
\end{eqnarray}
Substitution into Eq.\ (\ref{bilinear1}) then yields
\begin{eqnarray}
\label{4.14}
  F=\frac{\omega}{2\pi}\left(
\coth\frac{\hbar\omega}{2kT}-\frac{2kT}{\hbar\omega}\right).
\end{eqnarray}
In the opposite limit $\tau_0 \ll \hbar/kT$, $F(T,\omega,\tau_0)$ can be
found using the procedure proposed by Ivanov and Levitov \cite{IL}.
In a finite observation time $\tau_0$, energy levels are resolved 
up to $\hbar / \tau_0$, so that we must set
\begin{eqnarray}\label{n3}
f(t)=\frac{i}{2\tau_0}\frac{1}{\sin\left[\pi(t+i0)/
\tau_0\right]}.
\end{eqnarray}
Then, following Ref.\ \onlinecite{IL}, we change the time variable 
$t\in[0,\tau_0]$ to $z=e^{2\pi it/\tau_0}$, and perform contour
integration in the circle $|z|=1$ in complex plane. The result
coincides with the low-temperature limit of Eq.\ (\ref{4.14}) 
above.

The physical meaning of the function $F(T,\omega,\tau_0)$ with
$\tau_0 \gg 1/T$ becomes 
clear once it is written in energy representation,  
\begin{eqnarray}
  F &=&
  \int \frac{d\varepsilon}{2\pi\hbar}
  \left[ {1 \over 2} 
  f({\varepsilon+ \case{1}{2}\hbar\omega})
  {\tilde f}({\varepsilon- \case{1}{2}\hbar\omega})
  - f({\varepsilon}) {\tilde f}({\varepsilon}) 
  \right.  \nonumber \\ && \left. \mbox{}
  +  {1 \over 2} 
  f({\varepsilon- \case{1}{2}\hbar\omega})
  {\tilde f}({\varepsilon+ \case{1}{2}\hbar\omega})
  \right].
  \label{prob}
\end{eqnarray}
Equation (\ref{prob}) measures the change in the number of 
equilibrium electron-hole pairs due to absorption and emission of the 
pumping field quantum  $\hbar\omega$. 
At temperatures $kT\ll\hbar\omega$ the Fermi distribution is 
sharp, and $F=\omega/2\pi$. At high $kT\gg \hbar
\omega$, the Fermi distribution is smooth on 
the scale $\sim\omega$, so that $F$ is small, 
$F \sim \hbar\omega^2/kT$. 

Fluctuations of the noise are described by the second factor 
in Eq.\ (\ref{factor}), which varies from sample to sample. 
To find the mesoscopic fluctuations of the noise, we use 
the joint distribution of the matrices ${\cal R}_i$ ($i=1,2$) derived 
in Ref.\ \onlinecite{waves}. The resulting distribution of this mesoscopic
 contribution depends on two parameters,
\begin{eqnarray}\label{CcCl}
C_l&=&\frac 2N(X_1^2+X_2^2),\ \ C_c=\frac 4N X_1X_2\sin\phi.
\end{eqnarray}
Equality $C_c = C_l$ is achieved if $X_1=X_2$ and $\phi=\pi/2$, 
corresponding to the circular contour in the $(X_1,X_2)$ plane.  
Below we consider the noise distribution for a quantum dot with
single-channel point contacts ($N_{L}=N_{R}=1$), and for a dot with
many-channel point contacts ($N_{L},N_{R} \gg 1$) separately.


\subsubsection{Two channel geometry, $N=2$}\label{sec_211}

\begin{figure}
\centerline{\psfig{figure=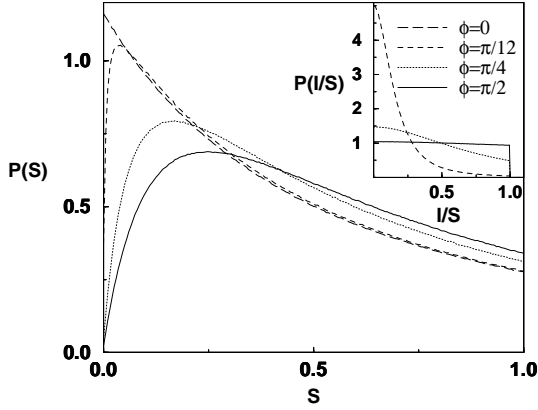,width=7.5cm}}
\caption{
Distribution of the pumping noise $S^P$ (main panel) and the
current-to-noise ratio $I/S^P$ (inset) for a quantum dot with two single
channel point contacts and broken time-reversal symmetry, for
various values of $C_c/C_l$. The values of $C_c/C_l$ shown are
$C_c/C_l = 1$, $\sin(\pi/4)$, $\sin(\pi/12)$, and $0$.
%
If the variations of the parameters $X_1$ and $X_2$
have equal amplitudes, this corresponds to phase difference
$\phi=\pi/2$, $\pi/4$, $\pi/12$, and $0$, respectively.
The noise $S^P$ is
measured in units of $e^2 C_l F $ and the current $I$ is measured
in units of $e C_l \omega/2 \pi$. For $C_c=0$, $P(I/S^P)$ is a
delta function at $I/S^P = 0$ (not shown).
}
\label{fig:015}
\end{figure}

For small $N$, the full distribution of the noise can be obtained
using the method of Ref.\ \onlinecite{trick} to numerically
generate the matrices ${\cal R}_1$ and ${\cal R}_2$ according to the
appropriate distribution, see Appendix for details. 
Figure \ref{fig:01} shows
the distribution of the noise power $S^P$ and the current-to-noise
ratio $I/S^P$ for a quantum dot
with single-channel leads ($N=2$) for the case $C_l=C_c$. 
The distributions
are shown with and without time-reversal symmetry (TRS). For
reference we have also included the distributions for the case
$e^2/C \ll \Delta$ of weak electron-electron interactions inside
the dot.
The case $C_c < C_l$ (i.e., the dependence
on the phase difference $\phi$) is illustrated in Fig.\
\ref{fig:015}. 
The current-to-noise ratio shown in Figs.\ \ref{fig:01} and
\ref{fig:015} accounts for the pumping noise $S^P$ only. The
Nyquist-Johnson noise $S^N$ presents a different noise source,
which will dominate over the pumping noise if $k T \gtrsim \hbar
\omega \sqrt{C_l}$.
\begin{figure}
\centerline{\psfig{figure=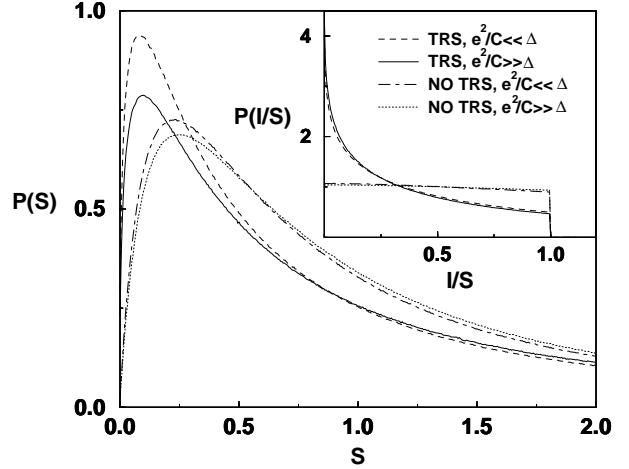,width=8.5cm}}
\caption{Main panel:
Distribution of pumping noise $S^P$ for a chaotic quantum
dot with single-channel point contacts 
and two time-dependent parameters $X_1$ and $X_2$ given by
Eq.\ (\protect\ref{eq:X}) with $C_l=C_c$. [The parameters
$C_l$ and $C_c$ are defined in Eq.\ (\protect\ref{CcCl}).]
The noise is measured in units of $e^2 C_l F $. The plots
are with and without time-reversal symmetry (TRS) and for
$e^2/C \ll \Delta$ (weak electron-electron interactions inside
the dot) or $e^2/C \gg \Delta$. Inset: Distribution of the
current-to-noise ratio $I/S^P$, measured in units of $\omega/(2 
\pi F e)$. 
There is no divergence at $I/S^P\to 0$.
}
\label{fig:01}
\end{figure}

We note that the
distributions of $S^P$ and $I/S^P$ are highly non-Gaussian.
In particular, the mean $\langle S^P \rangle$ of the
noise distribution is dominated by the algebraic tail for
large $S^P$, and is not representative of the distribution
itself. [For example, in the absence of time-reversal symmetry
and for $e^2/C \ll \Delta$,
the mean $\langle S^P \rangle = 8/3 (e^2 C_l F)$, while the
most probable value is for $S^P \approx 0.5 (e^2 C_l F)$.]
We also note that, while the phase difference $\phi$ affects the 
typical
size of the time-averaged pumped current $I \propto e \omega C_c$ 
but not the form of the distribution, changing
$\phi$ has a small effect on the average noise $\langle S^P \rangle$,
but changes the shape of the noise distribution significantly, see
Fig.\ \ref{fig:015}. In particular, the probability to find small
$S^P$ is significantly higher for $\phi$ close to zero than for
$\phi \sim \pi/2$. The reason for this difference is that the case
$\phi \sim \pi/2$ corresponds to noise generated by two independent 
sources, while $\phi$ close to zero corresponds to only a
single noise source. 
Furthermore, the current-noise ratio $I/S^P$ has a maximum at
$I/S^P = e$, as was predicted by
Levitov \cite{L}. For a quantum dot with single-channel point contacts,
there is a finite probability density to achieve this optimum
current-to-noise ratio, as is seen in the inset of Fig.\
\ref{fig:01}. 
For point contacts with more than one
channel, the probability density to attain the maximum value
$I/S^P=e$ vanishes \cite{L}, see, e.g. , Fig. \ref{fig:02}.

\subsubsection{Multichannel limit, $N\gg 1$}\label{sec_212}

For large $N$, the ensemble average and variance of the
noise can be expressed in terms of an integral over the unitary group
and over the eigenvalues of the Wigner-Smith time-delay matrix
${\cal R} = -i \hbar (\partial S/\partial \varepsilon) S^{\dagger}$ 
\cite{Smith}. These
integrals can be calculated using the method of Refs.\
\onlinecite{unitary,we}, together with asymptotic expressions
for the density and two-point correlations of the eigenvalues of the
Wigner-Smith time-delay matrix, see Appendix 
for details.

The result is
\begin{mathletters}\label{zerotnoise}
\begin{eqnarray}
\label{q_2weak}
  \langle S^P \rangle
   &=& 2 e^2 F g\left(1+\frac 2N\delta_{\beta,1}\right) C_l\\
  \mbox{var}\, S^P &=&  \left(2 e^2 F\right)^2\frac{g}{N}
\left(3\left[1+(1+2 \delta_{\beta,1})\frac {2g}{N}\right]C_l^2
\right.\nonumber \\ && \mbox{} \left.
-\left[1+(1+4\delta_{\beta,1})\frac gN\right]C_c^2\right)
\label{q_22weak}
\end{eqnarray}
\end{mathletters}
where we abbreviated $g = N_{L} N_{R}/N$.
The factor $F$ depends on $\omega$ and $T$ and was defined in Eq.\
(\ref{4.14}). 
Note that the fluctuations of $S^P$ are a factor $\sim 1/N$ smaller than
the average. Higher cumulants of the noise are
even smaller, so that we conclude that, for $N \gg 1$, the noise
distribution becomes sharply peaked at the average $\langle S^P
\rangle$ and that the remaining sample-to-sample fluctuations are Gaussian.

\begin{figure}
\centerline{\psfig{figure=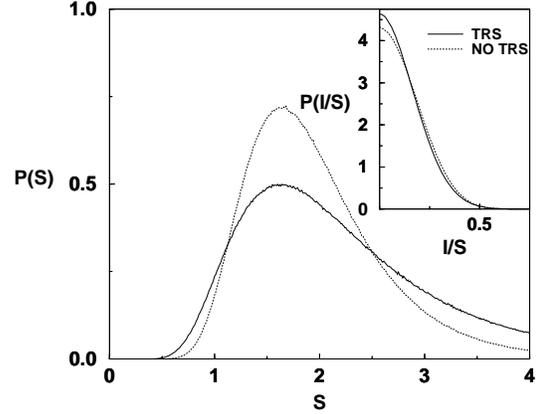,width=7.5cm}}
\caption{
The same as Fig. \ref{fig:01} for a quantum dot with $N_{L} = N_{R} = 5$.
Curves with weak and strong electron-electron interaction are
indistinguishable.
}
\label{fig:02}
\end{figure}
In Fig. \ref{fig:02} we show the results of a numerical calculation of 
the distribution
noise $S^P$ and the current-to-noise ratio $I/S^P$ for a dot with
$N_{L} = N_{R} = 5$, $N=10$. This value of $N$ can be seen as intermediate between
the small-$N$ regime, where the distributions are strongly
non-Gaussian and the large-$N$ regime where the distributions are
Gaussian. We note that for $N=10$ the noise distribution still has
pronounced tails for large $S^P$.

\subsection{Noise at arbitrary pumping for multichannel
 dots}\label{sec_22}

In this subsection we consider a general time dependence and 
amplitude of the parameters $X_j$. We limit
ourselves to the calculation of the ensemble averaged noise,
since the mesoscopic fluctuations of the noise
are smaller than the average by a factor $1/N$ if $N\gg 1$,
see Eq.~(\ref{zerotnoise}).

For a calculation of the ensemble average noise $\langle S^P \rangle$,
we need to know the correlation functions of scattering matrix elements
${\cal S}_{\alpha\beta}(t,t')$ for an ensemble of chaotic quantum
dots. The indices $\alpha$ and $\beta$ refer to the ``orbital''
channels as well as to the spin of the electrons. 
In order to discriminate between the orbital and spin degrees
of freedom, we set $\alpha = 
(i,\sigma)$, where $\sigma = \pm 1$ refers to spin and
$i=1,\ldots,N$ denotes the orbital channels, 
$i=1,\ldots,N_{L}$ for channels in the left point contact and 
$i=N_{L}+1,\ldots,N$ for channels in the right point contact.
In this notation, the scattering matrix ${\cal S}_{\alpha\beta}
= {\cal S}^{\rm o}_{ij} \delta_{\sigma \sigma'}$ is proportional
to the $2 \times 2$ unit matrix in spin space. The correlator of
the orbital part $S^{\rm o}_{ij}$ reads
\widetext
\Lrule
\begin{eqnarray}\label{corr}
\langle{\cal S}^{\rm o}_{ij}(t,t') 
{\cal S}_{kl}^{{\rm o}*}(\tau,\tau')\rangle &=&
\delta(t-t'-\tau+\tau')\left[ \delta_{ik}\delta_{jl}
\ {\cal D}(\frac{t+\tau}{2},t'+\frac{\tau-t}{2},\tau-t)+\delta_{il}
\delta_{jk}
\ {\cal C}(\frac{t+\tau}{2},t'+\frac{\tau-t}{2},\tau-t)\right],
\end{eqnarray}
with ${\cal D}$ and ${\cal C}$ given by
\begin{eqnarray}
\label{Dcorr}
{\cal D}(t_1,t_2,\tau)&=&\Theta(t_1-t_2)\exp\left\{- \!
  \int_{t_2}^{t_1}\! 
  {\Delta d\xi \over 2 \pi \hbar}
\left[N_d+
 \mbox{}
2\ ({\cal X}^{\rm T}({\xi-\tau/2})-{\cal X}^{\rm T}({\xi+\tau/2}))
({\cal X}({\xi-\tau/2})-{\cal X}({\xi+\tau/2}))\right]\right\}, \\
\label{Ccorr}
{\cal C}(t_1,t_2,\tau)& = &\Theta(t_1-t_2)\exp\left\{-\!
  \int_{t_2-\tau /2}^{t_1-\tau /2}\!   {\Delta d\eta \over 2 \pi \hbar}
\left[N_c+
2\ ({\cal X}^{\rm T}({\eta})-{\cal X}^{\rm T}({t_1+t_2-\eta}))
({\cal X}({\eta})-{\cal X}({t_1+t_2-\eta}))\right]\right\}.
\end{eqnarray}
\Rrule\narrowtext\noindent
Here $\Theta(z) = 1$ if $z > 0$ and $0$ otherwise and we abbreviated 
\begin{equation}
  N_d = N,\ \
  N_c = N + 2 x^2,
\end{equation}
where $x \propto (\Phi/\Phi_0)(\tau_{\rm erg} \Delta)^{-1/2}$ is a
dimensionless parameter describing the magnetic flux penetrating
the quantum dot and $\Phi_0$ the flux quantum \cite{BeenakkerReview}. 
The unitary ensemble, when time-reversal symmetry is fully broken,
corresponds to the limit $x \to \infty$.

In the literature, two equivalent approaches have been taken to
calculate correlators such as Eq.\ (\ref{corr}) above,
the Hamiltonian and scattering approaches \cite{BeenakkerReview}.
In the Hamiltonian approach the fundamental object is the random
Hamiltonian of the closed chaotic quantum dot and the Green functions 
related to it. Once the scattering matrix $\cal S$ is expressed in
terms of Green functions, the correlator (\ref{corr}) can be analyzed 
by standard diagrammatic techniques. The two terms
${\cal C}$ and ${\cal D}$ then appear as the cooperon
and diffuson contributions. 
The fundamental object of the scattering approach is a
statistical model for the
scattering matrix $\cal S$ of an ensemble of dots. Equivalence of
both methods, including the parametric and energy dependence of 
${\cal S}$ is shown in Ref.\ \onlinecite{waves}. A derivation of
Eq.\ (\ref{corr}) using the scattering approach is given in Ref.\
\onlinecite{future}. We refer to
Ref.\ \onlinecite{mvthesis} for a discussion based on the Hamiltonian
approach. 

Knowing the correlator (\ref{corr}), we can 
find the ensemble-averaged noise $\langle S^P \rangle$,
\widetext\Lrule
\begin{eqnarray}\label{diffusonnoise}
\langle S^P \rangle &=& \frac{2e^2 g}{\tau_0}\int_0^{\tau_0}dt\
dt'f(t-t')\tilde f(t'-t) 
\left(\left[\int_0^{\infty}{\cal D}(\frac{t+t'}{2},
\frac{t+t'}{2}-\frac{\hbar \zeta}{\esc},t'-t)\ d\zeta\right]^2-1\right),
\end{eqnarray}
\Rrule
\narrowtext\noindent
where $g = N_{L} N_{R}/N$ is the dot conductance, $\gamma=N\Delta/2\pi$ 
the escape rate, and we assumed $\tau_0 \gg
1/\omega, \hbar/\gamma$. The Cooperon contribution ${\cal C}$ of
Eq.\ (\ref{corr}) does not contribute to $\langle S^P \rangle$ to
leading order in $1/N$; its contribution is a factor $1/N$ smaller,
see, e.g., Eq.\ (\ref{q_2weak}).

Equation (\ref{diffusonnoise}) gives the ensemble-averaged
noise for arbitrary $\omega$, $T$, and $\gamma$, and for
arbitrary excursions of the parameters $X_j$.
We now investigate Eq.\ (\ref{diffusonnoise})
for the case that there are two time-dependent parameters
$X_1$ and $X_2$ with time dependence given by Eq.\
(\ref{eq:X}). 
In order to distinguish regimes of ``weak'' (bilinear)
and ``strong'' pumping, we introduce the energy scale 
\cite{VAA,transport}
\begin{eqnarray}\label{Tstar}
  k T^* = \hbar\omega \sqrt{C_l}.
\end{eqnarray}
The meaning of the energy scale $k T^*$ becomes clear when we
view the pumping process as ``diffusion in energy
space'': carriers absorb or emit energy quanta of size $\hbar \omega$ 
at a rate $X^2 \Delta/\hbar$ \cite{VAA}. 
Weak (bilinear) pumping corresponds to the case when the probability 
to absorb or emit one or more quanta is small, i.e., to the regime
$X^2 \Delta \ll \gamma$ or $k T^* \ll \hbar \omega$. 
For $k T^* \gg \hbar \omega$ many quanta are absorbed or emitted, so that
the carriers in the dot shift their energies by an amount 
$\sim \hbar \omega \sqrt{X^2 \Delta/\gamma} \sim T^*$. If $T^*$ exceeds
the temperature $T$ of the electrons in the leads, the time-dependent
potentials in the dot lead to significant ``heating'' of the electrons
inside the dot and $T^*$ can be viewed as an
effective electron temperature inside the dot. The latter
regime is referred to as ``strong'' pumping.

\begin{figure}
\centerline{\psfig{figure=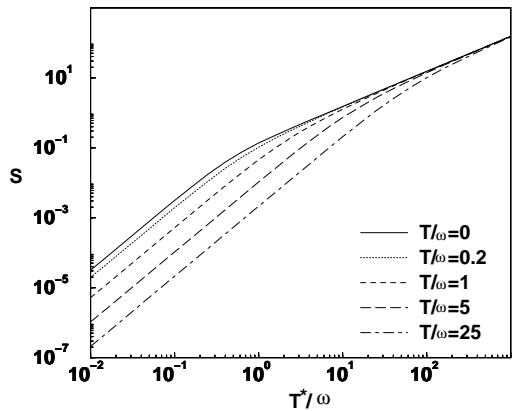,width=7.5cm}}
\caption{
Noise $\langle S^P \rangle$ in units of $e^2 g \omega$, as a function
of the dimensionless pumping strength $k T^*/\hbar \omega$ and
for various choices $k T/\hbar \omega$.
}\label{fig:0z}
\end{figure}

For weak pumping, $kT^*\ll\hbar \omega$, we can expand ${\cal D}$ to 
first order in $X^2$ and we recover
the result Eq. (\ref{q_2weak}), now without a restriction on 
the temperature $T$. For strong pumping, $k T^*\gg
\mbox{max}\{\hbar \omega, kT\}$, a simple expression
for the noise power can be obtained if pumping is adiabatic,
$\hbar \omega \ll \esc$. In that case, we note that
$\cal D$ in Eq. (\ref{diffusonnoise}) contains the fast decay
$\sim\exp(-\zeta)$ and a slowly varying contribution from the
time-dependence of the parameters $X_j$. Since the $X_j$ vary
slowly on the time scale $\hbar/\gamma$, the integration over 
$\zeta$ can be done, and the result is
\widetext\Lrule
\begin{eqnarray}\label{adiabatic}
 \langle S^P\rangle &=& 
\frac{2e^2 g}{\tau_0}\int_0^{\tau_0}dt'dtf(t'-t)\tilde f(t-t') 
  \left[\left(\frac{N}{N+2 ({\cal X}^{\rm T}(t)-{\cal X}^{\rm T}(t'))
  ({\cal X}(t)-{\cal X}(t'))}\right)^2-1\right]
\end{eqnarray}
In the limit of low temperatures, $k T\ll \hbar \omega$ (and, as
before, assuming long observation times
$\hbar/\tau_0 \ll kT, \hbar \omega$), Eq.\ (\ref{adiabatic}) yields
\begin{eqnarray}
  \label{T_0}
  \langle S^P \rangle & =&\frac{e^2g\omega}{\pi}
\frac{[1+6C_l(1+C_l(1-S^2))]\ {\bf E}(k)-[1+2C_l(1-S)]\ {\bf K}(k)}
{\pi[1+2C_l(1-S)]\sqrt{1+2C_l(1+S)}},
\end{eqnarray}
where ${\bf E}(k)$ and ${\bf K}(k)$ are full elliptic integrals of the 2nd
and 3rd kind, respectively, and we abbreviated
\begin{eqnarray}
S &=& \sqrt{1-\left(\frac{C_c}{C_l}\right)^2},\ \
k^2 = \frac{4C_l\ S}{1+2C_l(1+S)}.
\end{eqnarray}
In the special case that the two time-dependent parameters $X_1(t)$ and
$X_2(t)$ have equal amplitudes, $S = |\cos \phi|$.
The dependence of the averaged noise in Eq. (\ref{T_0}) on the
ratio $C_c/C_l$ (i.e., on the phase difference $\phi$) is weak. 
For the case $C_c = 0$ ($\phi=0$) we find the asymptotes
\begin{mathletters} \label{low}
\begin{eqnarray}\label{low1}
  \langle S^P \rangle
  &=& e^2 g {\omega \over \pi} C_l =
  e^2 g  {(k T^*)^2 \over \pi\hbar^2 \omega}\ \
  \mbox{if $k T^* \ll \hbar \omega$} \\
  \langle S^P \rangle
  &=& e^2 g {3 \omega \over \pi^2} \sqrt{\frac{C_l}{2}} =
  e^2 g {3 \over \pi^2\hbar\sqrt{2} } k T^* \ \
  \mbox{if $k T^* \gg \hbar \omega$}
  \label{low2}
\end{eqnarray}
\end{mathletters}
For the case $C_c=0$, but at arbitrary temperatures $kT$,
Eq. (\ref{adiabatic}) yields
\begin{eqnarray}\label{noisepower}
  \langle S^P \rangle
  &=& e^2 g\omega \left({2kT \over \hbar\omega}\right)^2
\int_0^\infty\frac{d\zeta}{\sinh^2(2\pi k T\zeta/\hbar \omega)}
\left\{
  1-\frac{\hbar \omega}{\sqrt{(\hbar \omega)^2+4 (k T^*)^{2} \sin^2 \zeta}}
  + {2 (k T^{*})^{2} \hbar \omega \sin^2 \zeta \over 
    \left[(\hbar \omega)^2+4 (kT^*)^{2}\sin^2 \zeta \right]^{3/2}} 
  \right\},
\end{eqnarray}
\Rrule\narrowtext\noindent
The low temperature asymptotics can be
easily obtained from this result, and reproduce 
the Eq.\ (\ref{low}) for $kT^* \ll \hbar\omega$ or 
$kT^* \gg \hbar\omega$. For intermediate values of $kT^*/\hbar\omega$ and 
$kT/\hbar \omega$, Eq.\ (\ref{noisepower}) is plotted in Fig.\ 
\ref{fig:0z}. As long as both $\hbar \omega$ and $k T$ are much
smaller than $k T^*$, the integral (\ref{noisepower}) is dominated
by $\zeta \ll \hbar \omega/k T^* \ll 1$, so that the strong pumping 
asymptote of Eq.\ (\ref{low2}) is reached, irrespective of 
$\omega$ or $T$.
An analytical expression for $\phi \neq 0$ can, in principle, 
be obtained
from Eq. (\ref{adiabatic}) as well. The qualitative behavior as a
function of pumping strength and temperature is similar to that
shown in Fig. \ref{fig:0z} for the case $\phi=0$. The limit of 
strong pumping should 
converge to the limit Eq.\ (\ref{T_0}) of $T=0$, similarly to the 
case $\phi = 0$ studied above.

\section{conclusion}\label{sec_3}

In this paper we calculated the current noise
generated in a quantum pump using a scattering matrix formalism with
a scattering matrix ${\cal S}(t,t')$ that depends on two times. With
this formalism, we could consider arbitrary pumping frequency $\omega$,
temperature $T$, escape rate $\gamma$, and pumping strength $X$. We
then calculated the average and variance of the noise for an ensemble
of quantum pumps consisting of a chaotic quantum dot. 

One issue that has received considerable attention recently is the
question whether one can build a noiseless quantum pump \cite{Avron2}. 
While our
results for the bilinear pumping regime show that there is a finite
(mesoscopic) probability for zero noise, it is not possible to have
no noise and a finite pumped current at the same time in the
bilinear regime
\cite{L,comment}, cf.\ Figs.\ \ref{fig:01} and \ref{fig:015}. This
is different beyond the bilinear regime, where a quantized, and,
hence, noiseless pump has been proposed using a pumping contour
that encircles a resonance in an almost closed quantum
dot \cite{Levinson1}.

We note that both the equilibrium Nyquist-Johnson noise $S^N$ and 
the pumping noise $S^P$ depend on the available energy 
window. For pumping noise, that energy window is the heating
temperature $T^*$, see Eq.\ (\ref{Tstar}); for Nyquist-Johnson
noise it is the temperature $T$ of the electron reservoirs.
The results (\ref{zerotnoise}) and
(\ref{T_0})--(\ref{noisepower})  allow us to compare the
Nyquist-Johnson and pumping contributions to the averaged noise. 
In the experimentally relevant case that $\hbar \omega \ll k T$,
both noise contributions are
proportional to the (dimensionless) dot conductance $g$, but the 
Nyquist-Johnson noise scales as $T$, while the pumping noise 
scales as $T^*$ if $T^* \gg T$
and as $T^{*2}/T$ if $T^*\ll T$. 
The Nyquist-Johnson
and pumping contributions to the noise are comparable at 
$T \sim T^*$. 
An experiment cannot separate the two contributions
to the noise, since it measures the total noise power. The
pumping contribution to the noise
is dominant as long as $T \ll T^*$.


\acknowledgements

We would like to thank Igor Aleiner for discussions. There is some
 overlap between our results and a recent preprint of Moskalets and 
B\"uttiker (Ref.\cite{buttikernew}), which appeared shortly before
 completion of our manuscript.
This work was supported by the Cornell Center for Materials 
Research under NSF grant No. DMR--0079992, by the NSF under
grants no.\ DMR--0086509 and DMR--0120702, and
by the Sloan and Packard foundations.

\appendix

\section*{ Integration over the matrices $\cal R$.}\label{appendix}

For the integration over the matrices ${\cal R}_1$ and ${\cal R}_2$
we make use of the fact that they can be parameterized as \cite{waves}
\begin{equation}
  {\cal R}_j = -i {\Delta \over 2 \pi \hbar}
  U \hat{\tau}^{1/2} H_j \hat{\tau}^{1/2} U^{\dagger} \otimes \openone_2,\ \ j=1,2,
\end{equation}
where $U$ is an $N \times N$ unitary matrix, $\hat{\tau}$ is a diagonal
$N \times N$ matrix containing the eigenvalues $\tau_m$,
$m=1,\ldots,N$, of the Wigner-Smith time-delay matrix on the
diagonal \cite{Smith}, $H_j$ is an $N \times N$ hermitian (real symmetric)
matrix is time-reversal symmetry is broken (present), and $\openone_2$
is the $2 \times 2$ unit matrix in spin grading.

For a chaotic quantum dot, the distributions of the
hermitian matrices $H_1$ and $H_2$, the unitary 
matrix $U$, and the diagonal matrix $\hat{\tau}$ are all independent.
The matrices $H_j$, $j=1,2$, have a Gaussian distribution,
\begin{equation}
  P(H) \propto \exp(-\beta \tr H^2/8),
\end{equation}
where $\beta=1$ if time-reversal symmetry is present and $\beta=2$
if time-reversal symmetry is broken by a magnetic field.
The matrix $U$ is uniformly
distributed in the unitary group, and the eigenvalues $\tau_m$,
$m=1,\ldots,N$ of the time-delay matrix have distribution 
\begin{eqnarray}
  P &\propto& 
  \left(1 + {e^2 \over \pi \hbar C} \sum_{m=1}^{N} \tau_m \right)
  \prod_{m<n}^{N} |\tau_m - \tau_n|^{\beta}
  \nonumber \\ && \mbox{} \times
  \prod_{m=1}^{N} \Theta(\tau_m) \tau_m^{-3 \beta N/2 + \beta - 2}
  e^{-\beta \pi \hbar/\Delta \tau_m}.
\end{eqnarray}

Knowing these distributions, finding the distribution for small
$N$ becomes a matter of mere quadrature. We have obtained the
plots of the distributions of $S^P$ and $I/S^P$ by numerically
generating $10^7$--$10^8$
matrices ${\cal R}_j$ distributed according to the
above distribution. We refer to Ref.\ \onlinecite{trick} for the
details of implementation of this procedure. Moments of the noise 
and the current can be found by performing the Gaussian integrations
over $H$ and the integrations over the unitary group with the help
of the technique of Ref.\ \onlinecite{unitary}. For small $N$, the
remaining integration over the $\tau_m$ can be done explicitly. For
large $N$, it is sufficient to know the density and two-point
correlator of the $\tau_m$ in order to find the first two moments
of $I$ or $S^P$. The density of time-delays is \cite{waves}
\begin{eqnarray}
\rho(\tau)&=& \sum_{m=1}^{N} \langle \delta(\tau_m-\tau) \rangle
=  \frac{N}{2\pi\tau^2}\sqrt{(\tau_{+}-\tau)
(\tau-\tau_{-})},\nonumber \\
\tau_{\pm}& =& 2 \pi \hbar ({3\pm\sqrt{8}})/{N\Delta}\nonumber
\end{eqnarray}
The pair correlation function $K_2(\tau_1,\tau_2)$ is a universal 
function of the arguments $\tau_1$ and $\tau_2$ and the ``spectrum
edges'' $\tau_-$ and $\tau_+$ \cite{brezinzee,BeenakkerUniv}. 
With the help of the pair correlation function \cite{BeenakkerUniv}
we find that, up
to corrections of order $1/N^4$,
\begin{eqnarray}
\left\langle\left(\sum_{m=1}^{N}\tau_m\right)^q\right\rangle &=& 
\left( {2 \pi \hbar \over \Delta} \right)^q
\left(1+q(q-1)\frac{2}{\beta N^2}\right).
\nonumber
\end{eqnarray}

\end{multicols}

\begin{references}
\bibitem{thouless}
D. Thouless, Phys. Rev. B {\bf 27}, 6083 (1983).
\bibitem{Niu} 
Q. Niu, Phys. Rev. Lett. {\bf 64}, 1812 (1990)

\bibitem{Kouwenhoven}
      L. P. Kouwenhoven, A. T. Johnson, N. C. van der Vaart,
      C. J. P. M. Harmans, and C. T. Foxon,
       Phys. Rev. Lett. {\bf 67}, 1626 (1991).
\bibitem{Pothier}
      H. Pothier, P. Lafarge, C. Urbina, D. Esteve, and M. H. Devoret,
       Europhys. Lett.\ {\bf 17}, 249 (1992).
\bibitem{Oosterkamp}
      T. H. Oosterkamp, L. P. Kouwenhoven, A. E. A. Koolen,
      N. C. van der Vaart, and C. J. P. M. Harmans,
      Phys. Rev. Lett. {\bf 78}, 1536 (1997).

\bibitem{FK} V. I. Fal'ko and D. E. Khmelnitskii, Zh. Eksp. 
Teor. Fiz. {\bf 95}, 328 (1989) [Sov. Phys. JETP {\bf 68}, 186
(1989)].

\bibitem{spivak} B. Spivak, F. Zhou, and M. T. Beal Monod, Phys. Rev. B
      {\bf 51}, 13226  (1995).


\bibitem{B_p} P. W. Brouwer, Phys. Rev. B {\bf 58}, 10135 (1998).

\bibitem{ZSA} F. Zhou, B. Spivak, and B. Altshuler, Phys. Rev. Lett.
{\bf 82}, 608 (1999).

\bibitem{Switkes} M. Switkes, C. M. Marcus, K. Campman, and A. C. Gossard,
Science {\bf 283}, 1905 (1999).

\bibitem{SAA} T. A. Shutenko, I. L. Aleiner, and B. L. Altshuler
Phys. Rev. B {\bf 61}, 10366 (2000).
\bibitem{Avron} J. E. Avron, A. Elgart, G.M. Graf, and L. Sadun,
  Phys. Rev. B {\bf 62}, 10618 (2000).

\bibitem{VAA} M. G. Vavilov, V. Ambegaokar, and I. L. Aleiner, Phys. Rev. B
  {\bf 63}, 195313 (2001).

\bibitem{Blaauboer}
M.\ Blaauboer and E.\ J.\ Heller, Phys. Rev. B {\bf 64}, 241301 (2001).

\bibitem{AK} A. V. Andreev and A. Kamenev,
  Phys.\ Rev.\ Lett.\ {\bf 85}, 1294 (2000).

\bibitem{L} L. S. Levitov, cond-mat/0103617
(2001)

\bibitem{AM} A. V. Andreev and E. G. Mishchenko, cond-mat/0104211.

\bibitem{LL} L. S. Levitov and G. B. Lesovik, Zh. Eksp. Teor. Fiz. {\bf 58},
 225 (1993) [JETP Lett. {\bf 58}, 230 (1993)].
\bibitem{LLL} L. S. Levitov, H.\ Lee, and G. B. Lesovik, J. Math. Phys. 
{\bf 37}, 4845 (1996).

\bibitem{IL} D. A. Ivanov and L. S. Levitov, 
  Pis'ma Zh. Eksp. Teor. Fiz. {\bf 58},
  450 (1993) [JETP Lett. {\bf 58}, 461 (1993)]; D. A. Ivanov, H.\ Lee and
  L. S. Levitov, Phys. Rev. B {\bf 56}, 6839 (1997).

\bibitem{MM} Yu.\ Makhlin and A. D. Mirlin, Phys. Rev. Lett. {\bf 87},
  276803 (2001).

\bibitem{buttikernew} M. Moskalets and M. B\"uttiker, cond-mat/0201259.

\bibitem{scatt} M. B\"uttiker, Phys. Rev. B {\bf 46}, 12485 (1992);
Phys. Rev. Lett, {\bf 65}, 2901 (1990);
Phys. Rev. B {\bf 45}, 3807 (1992).

\bibitem{buttiker1} M. B\"uttiker, A. Pr\^etre, and H. Thomas, Phys. 
Rev. Lett. {\bf 70}, 4114 (1993).
\bibitem{buttiker2} M. B\"uttiker, J. Phys. Condens. Matter {\bf 5},
9361 (1993); M. B\"uttiker, H. Thomas, and A. Pr\^etre, Z. Phys. B {\bf 94},
133 (1994).
\bibitem{BC}
  M. B\"uttiker and T. Christen, in {\em Quantum Transport in
      Semiconductor Submicron Structures}, B. Kramer ed., NATO ASI
      Ser.\ E, Vol.\ 326 (Kluwer, Dordrecht, 1996).
\bibitem{BB} Y. M. Blanter and 
  M. B\"uttiker, Phys. Rep. {\bf 336}, 1 (2001).

\bibitem{mvthesis} M. G. Vavilov, Ph.D. Thesis, Cornell University, (2001).

\bibitem{conductance}
M. G. Vavilov and I. L. Aleiner, Phys. Rev. B {\bf 60}, R16311 (1999);
{\bf 64}, 085115 (2001).

\bibitem{transport} V.\ I.\ Yudson, E.\ Kanzieper, and V.\ E.\ Kravtsov,
  Phys. Rev. B {\bf 64}, 045310 (2001).

\bibitem{BL} M.\ B\"uttiker and R.\ Landauer, Phys. Rev. Lett. {\bf
49}, 1739 (1982).

\bibitem{AppC}
  The first order correction in an expansion in powers of
  $\hbar\omega/\gamma$ is
  given in Appendix C of Ref.\ \onlinecite{VAA}. See also B.\ Wang,
 J.\ Wang, H.\ Guo, cond-mat/0107078 (2001), and O.\ Entin-Wohlman,
 A.\ Aharony, and Y.\ Levinson,  cond-mat/0201073 (2002)

\bibitem{parametric} P. W. Brouwer, S. A. van Langen, K. M. Frahm,
 M. B\"uttiker, and C. W. J. Beenakker, Phys. Rev. Lett. {\bf 79}, 913 (1997).

\bibitem{waves} P. W. Brouwer, K. M. Frahm, and C. W. J. Beenakker,
  Phys. Rev. Lett. {\bf 78}, 4737 (1997);
{\it Waves in Random Media} {\bf 9}, 91 (1999).

\bibitem{trick} J. N. H. J. Cremers and P. W. Brouwer, cond-mat/0110437 (2001)

\bibitem{unitary} P. W. Brouwer and C. W. J. Beenakker, J. Math. Phys
{\bf 37}, 4904 (1996).
\bibitem{we} M. L. Polianski and P. W. Brouwer, Phys. Rev. B. {\bf 64},
 075304 (2001).

\bibitem{Smith} E. P. Wigner, Phys. Rev. {\bf 98}, 145 (1955), and
F. T. Smith, Phys. Rev. {\bf 118}, 349 (1960).

\bibitem{BeenakkerReview}
C. W. J. Beenakker, Rev. Mod. Phys. {\bf 69}, 731 (1997).


\bibitem{future} M. L. Polianski and P. W. Brouwer, unpublished.

\bibitem{Avron2}
  J.E. Avron, A. Elgart, G.M. Graf, L. Sadun, Phys. Rev. Lett. {\bf 87}, 
  236601 (2001).

\bibitem{comment} A.\ Alekseev, cond-mat/0201474.

\bibitem{Levinson1} Y. Levinson, O. Entin-Wohlman, and P. W\"olfle,
  cond-mat/0002187.

\bibitem{brezinzee} 
E.\ Br\'ezin and A. Zee, Nucl. Phys. B {\bf 402}, 613 (1993).

\bibitem{BeenakkerUniv}
C. W. J. Beenakker, Nucl. Phys. B {\bf 422}, 515 (1994).

\end{references}
\end{document}